# Physical origin of color changes in lutetium hydride under pressure

Run Lv[1,2], Wenqian Tu[1,2], Dingfu Shao[1], Yuping Sun[3,1,4], Wenjian Lu[1, *]

[1]*Key Laboratory of Materials Physics, Institute of Solid State Physics, HFIPS, Chinese Academy of Sciences, Hefei 230031, China*
[2]*University of Science and Technology of China, Hefei 230026, China*
[3]*High Magnetic Field Laboratory, HFIPS, Chinese Academy of Sciences, Hefei 230031, China*
[4]*Collaborative Innovation Center of Microstructures, Nanjing University, Nanjing 210093, China*

Recently, near-ambient superconductivity was claimed in nitrogen-doped lutetium hydride ($LuH_{3-\delta}N_\varepsilon$) [1]. Unfortunately, all follow-up research still cannot find superconductivity signs in successfully synthesized lutetium dihydride ($LuH_2$) [2,3] and N-doped $LuH_{2\pm x}N_y$ [4-7]. However, a similar intriguing observation was the pressure-induced color changes (from blue to pink and subsequent red). The physical understanding of its origin and the correlation between the color, crystal structure, and chemical composition of Lu-H-N is still lacking. In this work, we theoretically studied the optical properties of $LuH_2$, $LuH_3$, and some potential N-doped compounds using the first-principles calculations by considering both interband and intraband contributions. Our results show that $LuH_2$ has an optical reflectivity peak around blue light up to 10 GPa. Under higher pressure, the reflectivity of red light gradually becomes dominant. This evolution is driven by changes in the direct band gap and the Fermi velocity of free electrons under pressure. In contrast, $LuH_3$ exhibits gray and no color change up to 50 GPa. Furthermore, we considered different types of N-doped $LuH_2$ and $LuH_3$. We found that N-doped $LuH_2$ with the substitution of a hydrogen atom at the tetrahedral position maintains the color change when the N-doping concentration is low. As the doping level increases, this trend becomes less obvious. While other N-doped structures do not show significant color change. Our results can clarify the origin of the experimental observed blue-to-red color change in lutetium hydride and also provide a further understanding of the potential N-doped lutetium dihydride.

## I. INTRODUCTION

The search for room-temperature superconductors has been a long-standing goal in condensed matter physics. According to BCS theory, metallic hydrogen and hydrogen-rich compounds with light ionic mass can produce high-temperature superconductivity [8]. In recent years, materials such as $H_3S$ (203 K at 155 GPa) [9,10] and $LaH_{10}$ (250 K at 190 GPa) [11,12] have been predicted by DFT calculations and confirmed by high-pressure experiments. Unfortunately, high pressure is always indispensable for the stability and superconductivity of these compounds [10]. Recently, near-ambient superconductivity was reported in N-doped Lu hydride $LuH_{3-\delta}N_\varepsilon$ [1]. The authors report that superconductivity emerges at 0.3 GPa, reaches its maximum at 1 GPa, and vanishes above 3 GPa. Interestingly, the crystal color changes from blue to pink at 0.3 GPa and subsequently to red at 3 GPa. However, the following research on lutetium dihydride $LuH_2$ [2,3] and $LuH_{2\pm x}N_y$ [4-7] show no near-ambient superconductivity but observed similar color changes under pressure. Other lutetium hydrides, such as $LuH_3$ [13] and $Lu_4H_{23}$ [14] show superconductivity at 12 K (122 GPa) and 71 K (218 GPa), respectively.

Undoubtedly, the microscopic crystal structure and chemical composition of lutetium hydride and its N-doped compounds are still controversial. However, the color change observed in several experiments is generally consistent. Color is a basic property of materials and determined by their optical reflectivity to the light with different wavelengths. Different structures and chemical compositions have different electronic properties, which determine their optical properties, such as reflection and absorption. Therefore, the color and its changes under pressure can provide basic information about the electronic structure of the material, which can be used as important evidence to study its structure and chemical composition. Despite numerous experimental reports on this system, the physical origin of the remarkable color changes under pressure remains a mystery. The difference between the optical properties of $LuH_2$ and $LuH_3$ is unclear. Furthermore, the possible occupation position and the effect of the doped nitrogen atom are still unknown.

In this work, we theoretically investigated the optical properties of $LuH_2$, $LuH_3$, and their potential N-doped compounds under pressure. Based on the dielectric function and electronic properties, we elucidate that both interband and intraband transitions are important for forming peaks of reflectivity spectra in the blue and red regions. Changes in its band gaps and Fermi velocity of free electrons are the origin of the remarkable color change. N-doped lutetium dihydride, $Lu_{18}H_{35}N$, with N replacing H at the tetrahedral site, exhibits a similar trend in reflectivity and blue-to-red color change as LuH$_2$ with increasing pressure. When the doping concentration of the N atom is higher, the color change becomes less obvious. However, the other five N-doped compounds considered in this work show no similar color change. The present results suggest that lutetium dihydride with N at the tetrahedral site is a potential nitrogen-doped lutetium hydride. Additionally, we calculated the phonon dispersion under different pressures and also considered anharmonic effects. The phonon dispersions indicate that LuH$_2$ is dynamically stable, while LuH$_3$ is unstable under near-ambient conditions.

---

* wjlu@issp.ac.cn

## II. CALCULATION DETAILS

The first-principles calculations based on Density Functional Theory (DFT) were performed by the Vienna Ab initio Simulation Package (VASP) [15,16]. The pseudopotential was described by using the projector augmented wave (PAW) methods and exchange-correlation interaction was treated by generalized gradient approximation (GGA), which is parameterized by Perdew-Burke-Ernzerhof (PBE) [17]. The plane-wave kinetic cutoff energy was set to 500 eV. Both lattice parameters and atom positions were fully optimized. The Brillouin zone was sampled with an 18×18×18 $k$-points mesh for structural optimization. The electronic properties calculation using a 40×40×40 $k$-points mesh for LuH$_2$ and LuH$_3$, at least a 10×10×10 $k$-points mesh for N-doped superstructures. Convergence criteria for energy and force were set to 10$^{-8}$ eV and 10$^{-3}$ eV/Å, respectively. The dielectric function (only for interband contribution) was calculated by the independent particle approximation (IPA) using the same $k$-points mesh with electron properties. For metals, the dielectric function has contributions from both interband and intraband transitions. The interband contribution describes all allowed direct transitions, defined as

$$\varepsilon^{inter}(\omega) = \varepsilon_1^{inter}(\omega) + i\varepsilon_2^{inter}(\omega).$$

It was post-processed with VASPKIT code [18]. The imaginary part of the 3 × 3 dielectric tensor is defined as [19]

$$\varepsilon_{2,\alpha\beta}^{inter}(\omega) = \frac{4\pi^2 e^2}{\Omega} \lim_{q \to 0} \frac{1}{q^2} \sum_{c,v,\mathbf{k}} 2w_k \, \delta(\epsilon_{c\mathbf{k}} - \epsilon_{v\mathbf{k}} - \omega) \times$$

$$\langle u_{c\mathbf{k}+\mathbf{e}_\alpha q} | u_{v\mathbf{k}} \rangle \langle u_{c\mathbf{k}+\mathbf{e}_\beta q} | u_{v\mathbf{k}} \rangle^*,$$

where the indices $c$ and $v$ denote the conduction and the valence band states, $w_k$ donates weight of $k$ points, $\mathbf{e}_\alpha$ and $\mathbf{e}_\beta$ are three Cartesian directions. The real part of the interband part is related to the imaginary part by the Kramers-Kronig relation

$$\varepsilon_{1,\alpha\beta}^{inter}(\omega) = 1 + \frac{2}{\pi} \mathbf{P} \int_0^\infty \frac{\varepsilon_{2,\alpha\beta}^{inter}(\omega')\omega'}{\omega'^2 - \omega^2} d\omega',$$

where P represents the principal value.

The intraband part was calculated by using the Drude model [20,21] as below,

$$\varepsilon^{intra}(\omega) = \varepsilon_1^{intra}(\omega) + i\varepsilon_2^{intra}(\omega),$$
$$\varepsilon_1^{intra}(\omega) = 1 - \frac{\omega_p^2}{\omega^2 + \Gamma^2},$$
$$\varepsilon_2^{intra}(\omega) = \frac{\omega_p^2 \Gamma}{\omega^3 + \omega\Gamma^2},$$

where $\omega_p$ and $\Gamma$ represent plasma frequency and relaxation rate, respectively. Here we select a typical value $\Gamma = 0.2$. And $\omega_p$ is obtained by

$$\omega_{p,\alpha\beta}^2 = \frac{4\pi e^2}{V\hbar^2} \sum_{n,k} 2g_k \frac{\partial f(E_{n,k})}{\partial E} (\mathbf{e}_\alpha \frac{\partial E_{n,k}}{\partial k})(\mathbf{e}_\beta \frac{\partial E_{n,k}}{\partial k}),$$

where $V$ donates the volume of the unit cell, $g_k$ is the weight of $k$ points, $f$ is the occupancy, $E$ is the band energy, and $\mathbf{e}_\alpha$ and $\mathbf{e}_\beta$ denote three directions. Finally, the total dielectric function obtained by

$$\varepsilon(\omega) = \varepsilon^{inter}(\omega) + \varepsilon^{intra}(\omega).$$

Then, the optical reflectivity is calculated by [22]

$$R(\omega) = \left| \frac{\sqrt{\varepsilon(\omega)} - 1}{\sqrt{\varepsilon(\omega)} + 1} \right|^2.$$

Colorimetry theory [23] is used to calculate the color of materials. Based on the reflectivity $R(\omega)$, the power distribution of reflected light under the D$_{65}$ standard illumination (simulates daylight) can be represented by

$$P(\omega) = P_{D_{65}}(\omega) \times R(\omega).$$

Then, the tristimulus values $(R, G. B)$, i.e., the color coordinates, are calculated by

$$R = \int_{380}^{780} P(\lambda) \bar{r}(\lambda) d\lambda,$$
$$G = \int_{380}^{780} P(\lambda) \bar{g}(\lambda) d\lambda,$$
$$B = \int_{380}^{780} P(\lambda) \bar{b}(\lambda) d\lambda.$$

The $\lambda$ represents the wavelength. The $\bar{r}(\omega)$, $\bar{g}(\omega)$ and $\bar{b}(\omega)$ are color-matching functions corresponding to the $(\mathbf{R}, \mathbf{G}, \mathbf{B})$ tristimulus space.

## III. RESULTS AND DISCUSSION

We first calculated the dielectric function of LuH$_2$ and LuH$_3$. The dielectric function reveals fundamental electronic structure information and determines optical properties, such as reflection and absorption. The color of the material can then be derived from the reflectivity. The dielectric function of interband transitions consists of a real part and an imaginary part. The imaginary part represents all allowed direct transitions from occupied to unoccupied states. The real and imaginary parts are related by the Kramers-Kronig relation [19]. Black lines in Figs. 1(b) and (d) show the interband dielectric function of LuH$_2$ under 0 GPa pressure. Considering that LuH$_2$ is a metal, intraband transitions also contribute to the dielectric function. The Drude free electron model [20,21] is used to describe the intraband part. The blue dashed lines in Figs. 1(b) and (d) represent the contribution of intraband transitions to the dielectric function. Compared to the dielectric function without intraband transitions, the inclusion of the intraband part significantly affects the imaginary dielectric function below 3 eV. Its effect on the real dielectric function is below 5 eV, due to the plasma frequency being 5.82 eV under ambient pressure. The calculated total dielectric function is also in good agreement with the experimental reports [24]. Figure 1(c) shows the reflectivity of LuH$_2$ under 0 GPa pressure. If only interband transitions are considered, the reflectivity exhibits little variation within the visible light range, as shown by the black line. However, when intraband transitions are taken into account, a peak appears in the blue light region and only a narrow energy range with high reflectivity near red light. Therefore, blue light is predominantly reflected, and the material has a blue appearance.

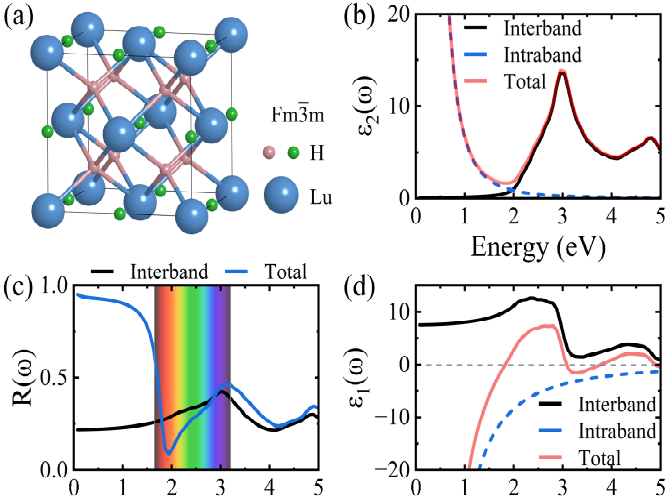

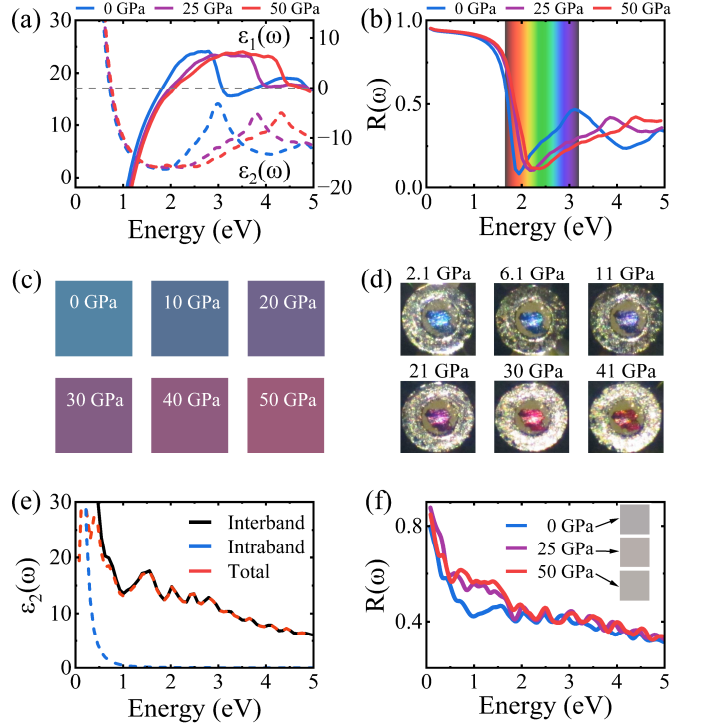

**Fig. 1.** (a) Crystal structure of lutetium hydride with the fluorite structure ($Fm\bar{3}m$). The pink and green H atoms occupy tetrahedral and octahedral sites, respectively. The structure of LuH$_3$ contains pink and green H atoms, while that of LuH$_2$ only contains pink H atoms. Calculated imaginary part $\varepsilon_2$ (b) and real part $\varepsilon_1$ (d) of the dielectric function of LuH$_2$. (c) Reflectivity $R(\omega)$ spectra of LuH$_2$.

To explore the pressure-induced color changes, we calculated the dielectric function and reflectivity under different pressures. Figure 2(a) shows the imaginary and real parts of the total dielectric function under different pressures. The plasma frequencies under different pressures are listed in Table S1. For the imaginary part $\varepsilon_2$ of the total dielectric function, its peak moves to higher energy when pressure increased, resulting in a decrease around 3 eV. This is mainly contributed by interband transitions. Below 2 eV, there is a slight increase, contributed by both interband and intraband transitions, as shown in Fig. 3(a). The real part $\varepsilon_1$ also shifts its peak to higher energy and decreases below 3 eV. These changes result in a variation of the reflectivity, as shown in Fig. 2(b). Under higher pressure, the reflectivity of blue light gradually decreases, while the reflectivity of red light increases. Consequently, the color of LuH$_2$ shifts to violet at ~20 GPa and to red at ~30 GPa. The reflectivity in 1.8 eV (red light) and 2.8 eV (blue light) under different pressures are listed in Table S1. The calculated color displayed in Fig. 2(c) is well consistent with the experimental observation [4,5] in Fig. 2(d). We also note the difference between the pressure of color transition calculated and observed in Refs. [2,3,7] could be mainly attributed to the presence of a small amount of N-doping and the different pressure conditions in the real experiments [7].

**Fig. 2.** (a) Imaginary part $\varepsilon_2$ (dashed line) and real part $\varepsilon_1$ (solid line) of the dielectric function of LuH$_2$ under different pressures. (b) Reflectivity spectra $R(\omega)$ of LuH$_2$. (c) Calculated color and (d) experimentally observed optical microscope images of LuH$_2$ under different pressures (from Ref. [4]) (e) Imaginary part of the dielectric function of LuH$_3$. (f) Reflectivity spectra $R(\omega)$ of LuH$_3$ under different pressures. Inset figures show the corresponding color.

To investigate whether LuH$_3$ could exhibit a similar color change, we calculated its dielectric function, reflectivity, and color under different pressures. Figure 2(e) shows the imaginary part of the dielectric function of LuH$_3$ under 0 GPa pressure. The plasma frequency of LuH$_3$ at 0 GPa pressure is 1.62 eV (the plasma frequencies at other pressures are also listed in Table S1). As a result, intraband transitions contribute to the dielectric function of LuH$_3$ mainly below 1 eV, which is lower than the energy range of visible light. Therefore, the total dielectric function in the visible light energy range is mainly determined by interband transitions. Moreover, the dielectric function within the energy range of visible light shows slight variation and does not have a prominent peak, resulting in a small variation of reflectivity in this range. Using its reflectivity, we calculated the color of $LuH_3$ under different pressures and found that it remains gray, as displayed in the inset of Fig. 2(d).

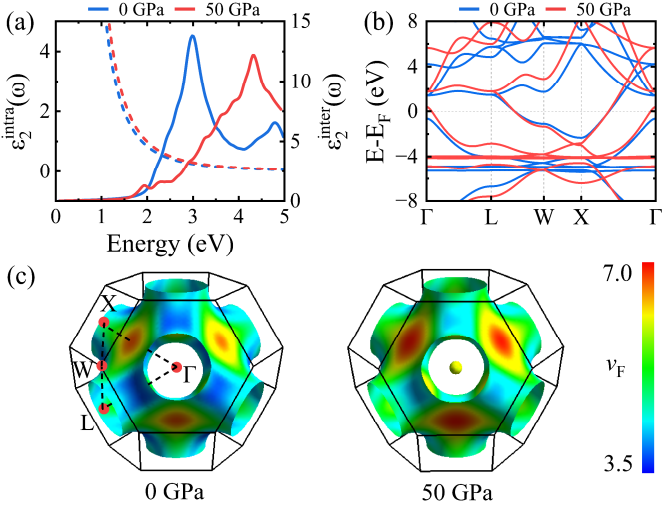

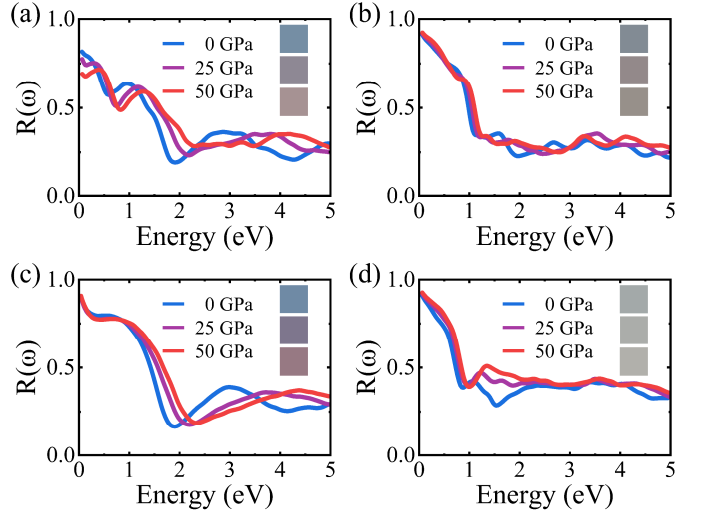

**Fig. 3.** (a) Intraband (dashed line) and interband (solid line) imaginary part of the dielectric function, (b) band structure and (c) Fermi surface of $LuH_2$ under 0 and 50 GPa.

**Fig. 4.** Reflectivity of (a) $Lu_8H_{15}N$ (nitrogen atom at the tetrahedral site), (b) $Lu_8H_{16}N$ (nitrogen atom at the octahedral site), (c) $Lu_{18}H_{35}N$ (nitrogen atom at the tetrahedral site), (d) $Lu_8H_{23}N$ (nitrogen atom at the octahedral site). Insets show the color at the corresponding pressure.

To gain a deeper understanding of the origin of the change in optical properties, we analyze the electronic structure of $LuH_2$. In Fig. 2(a), the imaginary part of the total dielectric function increases around 1.8 eV and decreases around 2.8 eV, resulting in color changes. These changes can be attributed to interband and intraband transitions, as shown in Fig. 3(a). Compared to the blue solid line (0 GPa), the peak of the red solid line (50 GPa) moves to 4.3 eV, resulting in a decrease of reflectivity in 2.8 eV. And around 1.8 eV, the red line shows a slight increase. In its band structure, the bandgap at the $W$ point increased from 2.8 eV to 4.3 eV, while the bandgap at the $\Gamma$ point decreased from 2.0 eV to 1.8 eV. These changes correspond to the variations in interband transitions. The red dashed line represents the intraband part, and it also shows a slight increase in 1.8 eV. This is due to the increase in the Fermi velocity of electrons on the Fermi surface, increasing the $\omega_p$ (listed in Table S1). Figure 3(c) shows the Fermi surface of $LuH_2$ at 0 GPa and 50 GPa, with red indicating regions for high Fermi velocity. The increase in Fermi velocity is accompanied by a decrease in effective mass, which results in higher electron mobility. This causes $LuH_2$ to exhibit better metallic behavior at higher pressures [5].

Furthermore, we considered the effect of different N doping on the color of Lu-H-N compounds. As shown in Fig. 1(a), there are two types of H atom positions in the $Fm\overline{3}m$ structure of lutetium hydride: tetrahedral and octahedral sites. For $LuH_2$, nitrogen can replace an H atom at the tetrahedral site or occupy the octahedral site. Firstly, we simply considered doping one N atom into a 2×2×2 supercell of $LuH_2$, which can form $Lu_8H_{15}N$ and $Lu_8H_{16}N$. We calculated their dielectric functions, reflectivity, and color. Their reflectivity and color under different pressure are shown in Figs. 4(a) and (b). Compared to $LuH_2$, there was no significant color change from blue to red. However, the reflectivity of $Lu_8H_{15}N$ in the visible light range exhibits a similar distribution to that of $LuH_2$. Under high pressure, the reflectivity of blue light decreases while that of red light increases. In contrast, the reflectivity of $Lu_8H_{16}N$ does not exhibit the above characteristics. This similar trend in reflectivity of $Lu_8H_{15}N$ under high pressure inspired us to consider that, at lower N doping concentrations, it may be possible to exhibit a significant color change similar to $LuH_2$. Therefore, we considered replacing one H atom with an N atom in a 3×3×2 supercell of $LuH_2$, forming $Lu_{18}H_{35}N$. Figure 4(c) shows that $Lu_{18}H_{35}N$ exhibits a very similar trend of reflectivity and color change under different pressures, suggesting $Lu_{18}H_{35}N$ is a potential structure of nitrogen-doped lutetium dihydride.

Next, we adopted four structures for N-doped $LuH_3$ that were previously suggested in Ref. [25]. These compounds have a high density of states near the Fermi energy, which can benefit the occurrence of superconductivity [25]. However, the calculated colors of the four compounds do not exhibit the blue-to-red color change. Figure 4(d) shows the results of $Lu_8H_{23}N$, where one H atom is replaced by an N atom at the octahedral site in a 2×2×2 supercell of $LuH_3$. Although the reflectivity of red light increases with increasing pressure, the change is small and the reflectivity of other colors in the visible light range is similar, resulting in a gray color under different pressures. Figure S5 shows the reflectivity of $Lu_8H_{23}N$ with N at the tetrahedral site. It shows a similar distribution to that of

LuH$_2$, but it does not have a similar evolution under pressure. Thus, Lu$_8$H$_{23}$N with N at the tetrahedral site preserves its blue color under different pressures. The other two compounds also exhibit gray color under different pressures (see Figs. S6 and S7). All considered structures and related dielectric functions of the N-doped lutetium hydrides are displayed in the supplementary materials.

Additionally, we calculated the phonon dispersion of LuH$_2$ and LuH$_3$ to investigate their structural stability. Figures S8(a) and (b) show that LuH$_2$ is dynamically stable from 0 GPa to 50 GPa, while LuH$_3$ is unstable at low pressures (< 25 GPa). For LuH$_3$, as pressure is higher than 25 GPa, the imaginary frequency vanishes. Using the mode decomposition technique developed in DynaPhoPy code [26], we also calculated the anharmonic phonon dispersion at 300 K. The comparison with harmonic and anharmonic phonon dispersion is shown in Figs. S8(c) and (d). When anharmonic effects are included, the phonon dispersion of LuH$_2$ shows only minor corrections. For LuH$_3$, most of the imaginary frequencies have vanished, but large imaginary frequencies remain near the $W$ point, indicating its instability. The above results are consistent with the previous report on the thermodynamic stability of lutetium hydride [27].

## IV. CONCLUSION

In summary, we theoretically investigated the optical properties of LuH$_2$, LuH$_3$, and different nitrogen-doped lutetium hydrides under different pressures. By calculating the dielectric function and reflectivity, we found that LuH$_2$ can reproduce the pressure-induced color changes observed in experiments. Both interband and intraband transitions contribute to the formation of reflectivity peaks in the blue-light and red-light regions. Under high pressure, the increase in the direct energy gap at the $W$-point, the decrease in the gap at the $\Gamma$-point, and the increase in the Fermi velocity of free electrons cause the reflectivity of blue light to decrease and that of red light to increase, resulting in the blue-to-red color change. Furthermore, the results of different N-doped compounds show only lutetium dihydride, with N replacing H at the tetrahedral site, could exhibit a similar trend of color changes of LuH$_2$ when the doping concentration of N atom is low. Other N-doped compounds show no significant color change. Additionally, our calculated phonon dispersions show that LuH$_3$ is dynamically unstable under near-ambient conditions. Our results clarify the physical origin of the pressure-induced blue-to-red color change of lutetium hydride observed in experiments and also provide a further understanding of the potential N-doped lutetium dihydride.

## ACKNOWLEDGEMENT

This work was supported by the National Key Research and Development Program of China under Contract No. 2022YFA1403203 and No. 2021YFA1600200, the National Nature Science Foundation of China under Contract No. U2032215 and No. 12241405.

# Supplemental Material

# Physical origin of color changes in lutetium hydride under pressure

Run Lv[1,2], Wenqian Tu[1,2], Dingfu Shao[1], Yuping Sun[3,1,4], Wenjian Lu[1, *]

[1]*Key Laboratory of Materials Physics, Institute of Solid State Physics, HFIPS, Chinese Academy of Sciences, Hefei 230031, China*
[2]*University of Science and Technology of China, Hefei 230026, China*
[3]*High Magnetic Field Laboratory, HFIPS, Chinese Academy of Sciences, Hefei 230031, China*
[4]*Collaborative Innovation Center of Microstructures, Nanjing University, Nanjing 210093, China*

Table S1 lists the plasma frequencies and reflectivity of blue and red light of $LuH_2$ and $LuH_3$ under different pressures.

Figures S1-S7 show crystal structures and dielectric functions of N-doped lutetium hydrides calculated in this work. The contributions of interband and intraband transitions to dielectric functions at 0 GPa are plotted separately, and the evolution of the total dielectric function under different pressures is also shown. The reflectivity and color of those structures not displayed in the main text are shown here.

Figure S8 shows the calculated phonon dispersion of $LuH_2$ and $LuH_3$. Figs. S8(a) and (b) show the harmonic phonon dispersion. It was calculated using the density-functional perturbation theory (DFPT) method, with phonon frequency determined via the PHONOPY code [1]. Figs. S8(c) and (d) show the anharmonic phonon dispersion. The anharmonic phonon frequency is calculated using the mode decomposition technique developed in DynaPhoPy [2]. This code uses the molecular dynamics trajectory calculated by VASP and projects it onto a set of harmonic phonon modes obtained by Phonopy. The power spectrum is then calculated using the maximum entropy method (MEM), and the peaks are fitted to Lorentzian functions to extract the anharmonic phonon properties.

**Table S1**. The plasma frequencies and reflectivity of typical blue (2.8 eV) $R_{blue}$ and red (1.8 eV) $R_{red}$ energy range under different pressures.

| Structure | $LuH_2$ | | | $LuH_3$ | | |
|---|---|---|---|---|---|---|
| Pressure | $\omega_p$ (eV) | $R_{blue}$ | $R_{red}$ | $\omega_p$ (eV) | $R_{blue}$ | $R_{red}$ |
| 0 GPa | 5.82 | 0.45 | 0.28 | 1.62 | 0.42 | 0.41 |
| 10 GPa | 6.00 | 0.34 | 0.25 | 2.18 | 0.42 | 0.41 |
| 20 GPa | 6.11 | 0.29 | 0.48 | 2.36 | 0.44 | 0.41 |
| 30 GPa | 6.17 | 0.27 | 0.57 | 2.89 | 0.44 | 0.42 |
| 40 GPa | 6.25 | 0.25 | 0.60 | 2.74 | 0.43 | 0.44 |
| 50 GPa | 6.33 | 0.23 | 0.60 | 2.22 | 0.44 | 0.43 |

---

* wjlu@issp.ac.cn

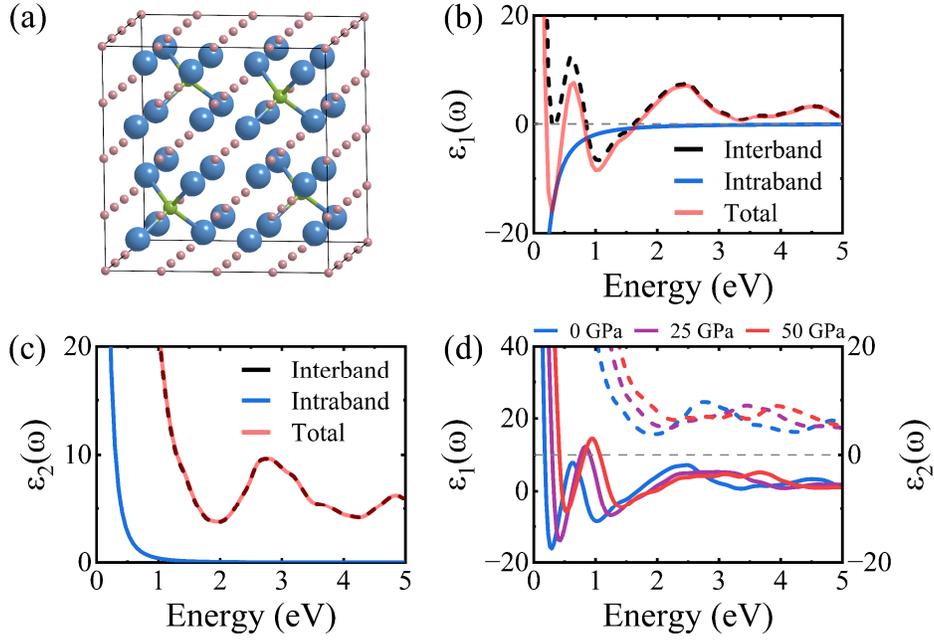

**Fig. S1.** Crystal structure (a) and dielectric functions (b)-(d) of $Lu_8H_{15}N$ (nitrogen atom at the tetrahedral site). Solid lines and dashed lines in (d) represent the real part $\varepsilon_1$ and imaginary part $\varepsilon_2$ of the dielectric function, respectively.

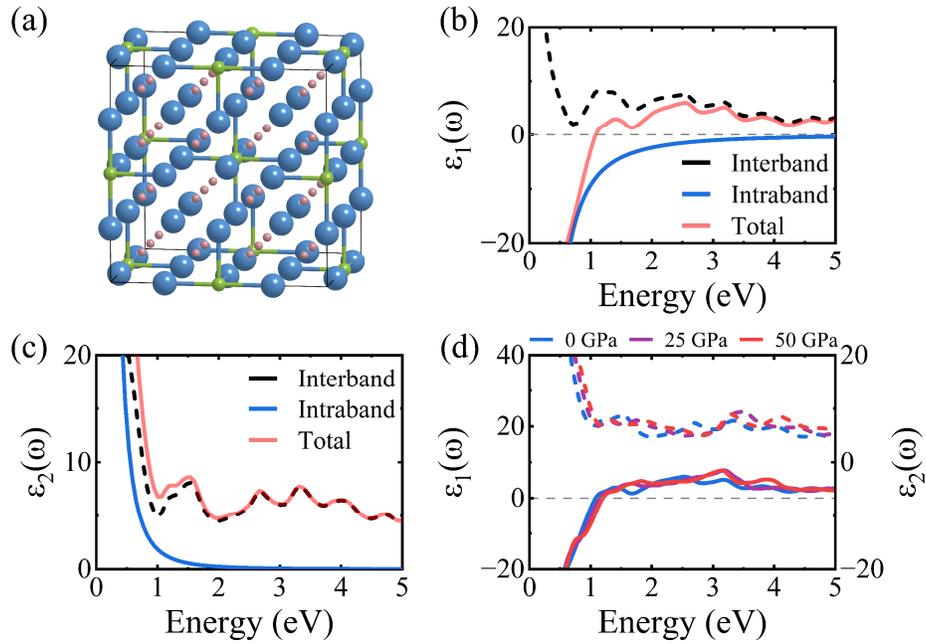

**Fig. S2.** Crystal structure (a) and dielectric functions (b)-(d) of $Lu_8H_{16}N$ (nitrogen atom at the octahedral site).

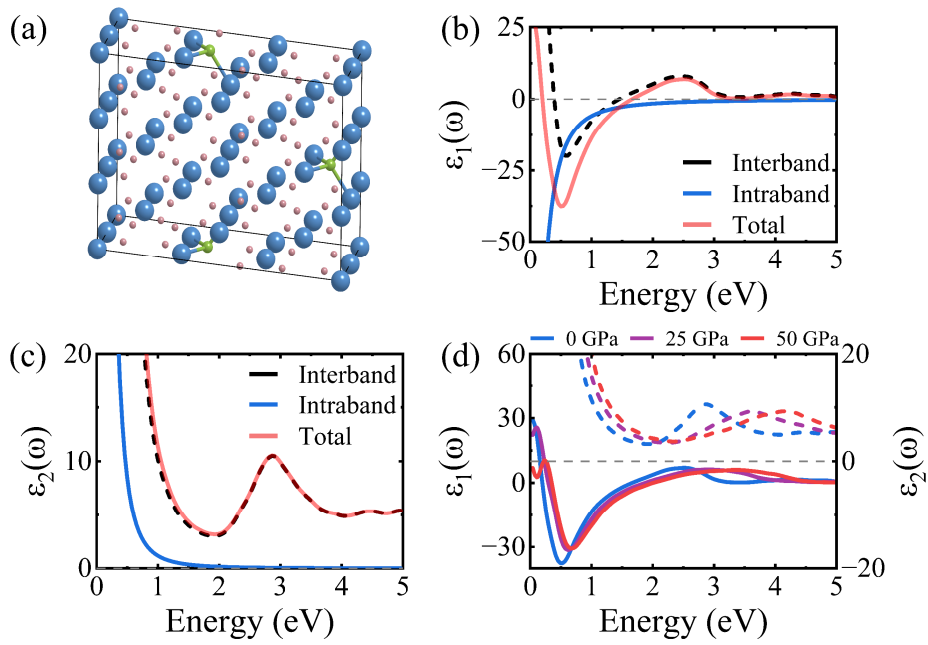

**Fig. S3.** Crystal structure (a) and dielectric functions (b)-(d) of $Lu_{18}H_{35}N$ (nitrogen atom at the tetrahedral site).

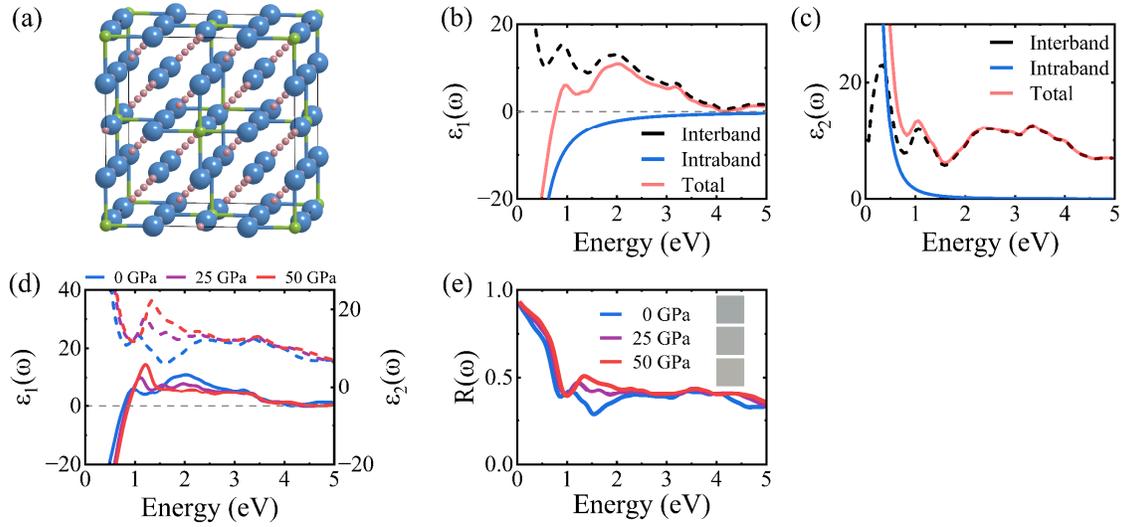

**Fig. S4.** Crystal structure (a) and dielectric functions (b)-(d) of $Lu_8H_{21}N$ (nitrogen atom at the octahedral site).

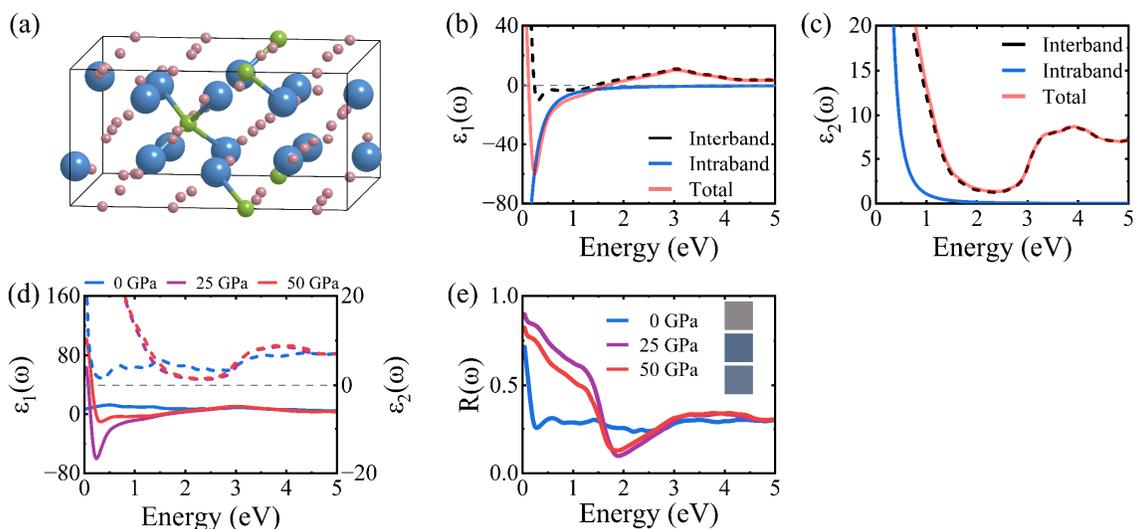

**Fig. S5.** Crystal structure (a) and dielectric functions (b)-(d), (e) reflectivity and color of $Lu_8H_{23}N$ (nitrogen atom at the tetrahedral site).

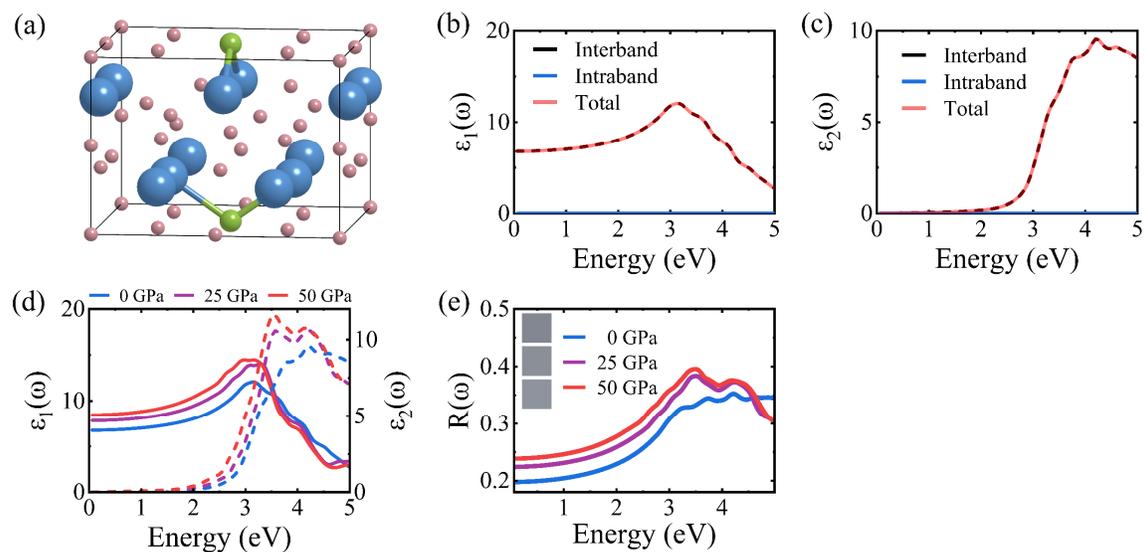

**Fig. S6.** (a) Crystal structure, (b)-(d) dielectric functions, (e) reflectivity and color of $Lu_8H_{21}N$ (nitrogen atom at the tetrahedral site).

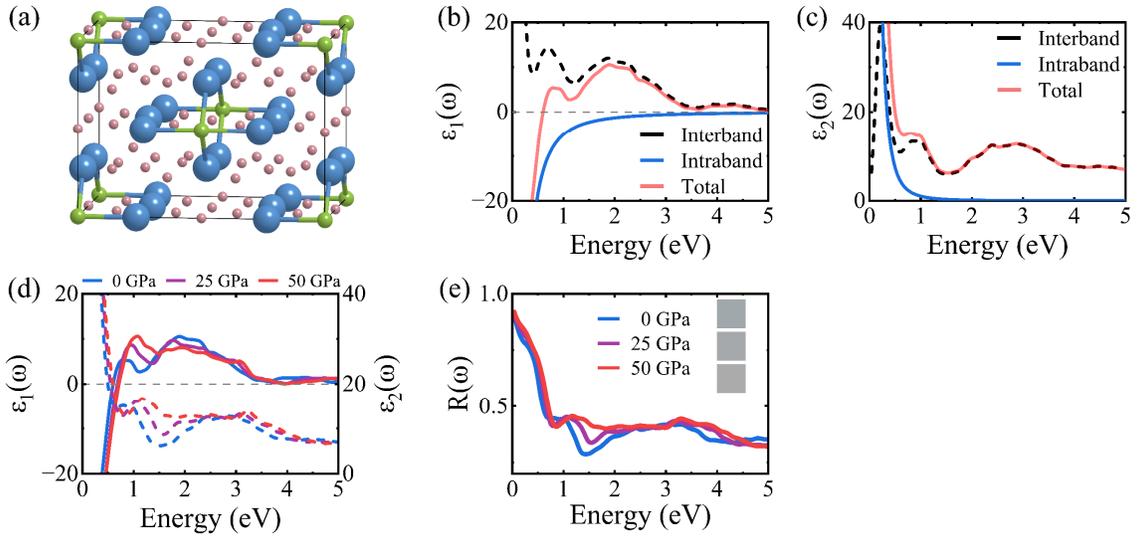

**Fig. S7.** (a) Crystal structure, (b)-(d) dielectric functions, (e) reflectivity and color of $Lu_8H_{23}N$ (nitrogen atom at the octahedral site).

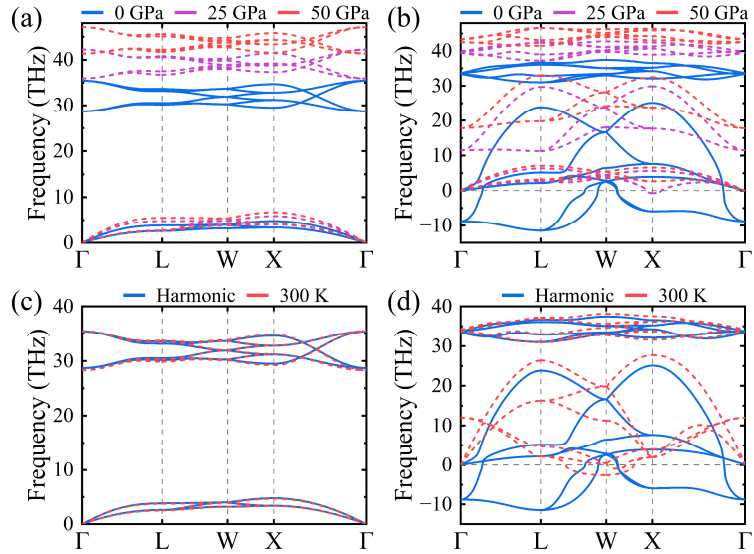

**Fig. S8.** Harmonic phonon dispersion of (a) $LuH_2$ and (b) $LuH_3$ under different pressures. Harmonic and anharmonic phonon dispersion at 300 K of (c) $LuH_2$ and (d) $LuH_3$.